\documentclass[reprint,amsmath,amssymb,jap,aip,twocolumn]{revtex4-1}
\usepackage{graphicx}
\usepackage{dcolumn}
\usepackage{bm}
\usepackage{epsfig}
\usepackage{mathptmx}
\usepackage{times}
\usepackage{amsmath}
\usepackage{amssymb}
\usepackage{units}
\usepackage{upgreek}
\usepackage{float}
\usepackage{color}
\usepackage{soul}

\begin{document}
\title{Mixed-precision training of deep neural networks using computational memory}
\author{Nandakumar S. R.}\email{nsi@zurich.ibm.com}
\affiliation{IBM Research -- Zurich, 8803 R\"{u}schlikon, Switzerland.}\affiliation{New Jersey Institute of Technology (NJIT), Newark, NJ 07102, USA}
\author{Manuel Le Gallo}
\affiliation{IBM Research -- Zurich, 8803 R\"{u}schlikon, Switzerland.}
\author{Irem Boybat}
\affiliation{IBM Research -- Zurich, 8803 R\"{u}schlikon, Switzerland.}\affiliation{Ecole Polytechnique Federale de Lausanne (EPFL), 1015 Lausanne, Switzerland}
\author{Bipin Rajendran}
\affiliation{New Jersey Institute of Technology (NJIT), Newark, NJ 07102, USA}
\author{Abu Sebastian}\email{ase@zurich.ibm.com}
\affiliation{IBM Research -- Zurich, 8803 R\"{u}schlikon, Switzerland.}
\author{Evangelos Eleftheriou}
\affiliation{IBM Research -- Zurich, 8803 R\"{u}schlikon, Switzerland.}
\date{\today}

\begin{abstract}
Deep neural networks have revolutionized the field of machine learning by providing unprecedented human-like performance in solving many real-world
problems such as image and speech recognition. Training of large DNNs, however, is a computationally intensive task, and this necessitates the
development of novel computing architectures targeting this application. A computational memory unit where resistive memory devices are organized in
crossbar arrays can be used to locally store the synaptic weights in their conductance states. The expensive multiply accumulate operations can be
performed in place using Kirchhoff's circuit laws in a non-von Neumann manner. However, a key challenge remains the inability to alter the
conductance states of the devices in a reliable manner during the weight update process. We propose a mixed-precision architecture that combines a
computational memory unit storing the synaptic weights with a digital processing unit and an additional memory unit accumulating weight updates in
high precision. The new architecture delivers classification accuracies comparable to those of floating-point implementations without being
constrained by challenges associated with the non-ideal weight update characteristics of emerging resistive memories. A two layer neural network in
which the computational memory unit is realized using non-linear stochastic models of phase-change memory devices achieves a test accuracy of 97.40\%
on the MNIST handwritten digit classification problem.
\end{abstract}
\maketitle

\section{Introduction}
Deep neural networks (DNN) including multilayer perceptrons, convolutional neural networks, deep belief networks, and Long-Short-Term-Memories are
loosely inspired by biological neural networks in which parallel processing units called neurons are interconnected by plastic synapses. By tuning the weights of the interconnections these networks are able to solve problems which are intractable by conventional algorithms. Through a combination of factors,  such as the availability of massive labeled datasets and the highly parallel matrix manipulations
offered by modern GPUs, these networks have recently achieved considerable success in numerous applications\cite{Y2015lecunNature}.

A DNN comprises multiple layers of neurons interconnected by synapses. Training of DNNs refers to the process of finding  appropriate synaptic weights such that after the training process, the network is able to perform various classification tasks with sufficient accuracy. Typically, this is achieved by a supervised training algorithm known as backpropagation. During the training phase, the input data is forward-propagated through the neuron layers with the synaptic network performing a multiply-accumulate operation. The final layer responses are compared with input data labels and the errors are back-propagated. All the synaptic weights are updated to reduce the error. This forward-backward data propagations and weight updates are repeated several times over the entire training data set. This brute force approach to training neural networks is computationally intense and, in spite of the availability of computing resources such as the GPUs, is very time-consuming. Also, the high power consumption of this training approach makes its application  prohibitive in several emerging domains such as internet of things and edge computing. Much of the inefficiency arises from the fact that the DNNs are trained using conventional von Neumann computing systems where the physical separation between the memory and processing units leads to constant shuttling of data back-and-forth between them.

Recently, there is a significant interest in designing non-von Neumann co-processors for training DNNs. A system comprising dense crossbar arrays of
resistive memory devices has been proposed to perform the various steps involved in the training of DNNs
\cite{Y2015burrTED,Y2016gokmenFN,Y2017burrAPX}. The devices, also referred to as memristive devices, store information in their conductance states
\cite{Y2011chuaAPA,Y2015wongNatureNano} and can be used to represent the synaptic weights. The matrix-vector multiplications needed during
the propagation of data in the network can then be computed as a result of Kirchhoff's circuit laws. Weight updates can be applied by modifying the
conductance levels of the resistive memory devices by applying appropriate electrical pulses. However, this approach can achieve
satisfactory training accuracy only with ideal, not-yet-available resistive memory devices \cite{Y2016gokmenFN}. The experimental demonstrations
based on existing resistive memory devices have achieved reduced classification accuracies owing to the inability to achieve precise conductance changes
in the memristive devices \cite{Y2015burrTED,Y2017boybatPRIME}.

A related research area that is gaining a lot of traction is in-memory computing or computational memory. Here,  physical attributes of the memory devices are exploited to perform computations in a non-von Neumann manner. There are recent demonstrations of performing bulk bit-wise operations\cite{Y2016seshadriArXiv}, matrix-vector multiplications\cite{Y2016huDAC,Y2017sheridanNatNano,Y2017legalloIEDM,  Y2017legalloArXiv} and finding temporal correlations using such a memory unit\cite{Y2017sebastianNatComm}. One major issue in this field is the limited precision of the individual units of the computational memory. Recently, we proposed the concept of mixed-precision in-memory computing to counter this challenge\cite{Y2017legalloArXiv}. The essential idea here is to use the low-precision computational memory unit in conjunction with a precise computing unit. The benefits of areal/energy/speed improvements arising from computational memory are retained while addressing the key challenge of inexactness associated with computational memory. As an example, we presented an iterative solver for systems of linear equations.

Meanwhile, there are some key developments taking place at the algorithmic front
with respect to training DNNs using digital arithmetic with reduced precision \cite{Y2015guptaICML,Y2015mullerArXiv,Y2015courbariauxANIPS,merolla2016,Zhang2017}. Recent work
shows that it is possible to have binary precision for the
weights used in the multiply-accumulate operations (during the
forward and backward propagations) as long as the precision
of the stored weights in which gradients are accumulated is
retained \cite{Y2015courbariauxANIPS}. This indicates the possibility of
accelerating the DNN training using programmable low precision computational memory, provided we address the challenge of reliably transferring the
high precision gradient to them.

In this article, we present a mixed-precision architecture based on computational memory to train DNNs. We investigate various undesirable attributes
of the constituent devices in  such a computational memory unit and show how the proposed architecture is designed to cope with them.
Finally, the DNN training performance of the mixed-precision scheme is evaluated where computational memory devices are realized using stochastic models based on  $90\,$nm phase-change memory characterization.
\section{Computational memory: the key challenges}

Non-volatile resistive memory devices have several attributes making them suitable candidates for building computational memory elements. These devices operate based on a variety of physical
mechanisms such as field driven atomic rearrangement (metal-oxide based resistive memory \cite{Y2012wongIEEEProc} (ReRAM) and conductive bridge
memory (CBRAM)\cite{Y2011valovNanotechnology}), spintronic effects (spin transfer torque based magnetic memory (STT-MRAM)\cite{Y2015kentNatureNano}
and phase transition (phase-change memory (PCM)\cite{Y2010wongIEEEProc}).  Irrespective of the underlying physical mechanism, all these devices store
information in their resistance or conductance states which are programmed by the application of suitable electrical pulses.  However, there are many
challenges associated with programming a desired conductance change in these devices. First, there are limitations on the minimum conductance change
that can be reliably induced. For example in STT-MRAM, it is very difficult to achieve more than two conductance levels due to the underlying
physical mechanism. Similarly in filamentary resistive memory devices such as CBRAM, the positive feedback mechanism involved in the filament growth
process makes it difficult to control the process and to achieve intermediate states\cite{Y2016nandakumarNanoletters}. This inability to achieve a
sufficiently small conductance change also limits the storage resolution. Another major challenge arises from stochasticity associated with the
device programming. In these nanoscale devices, slight changes in atomic configurations can lead to significantly different conductance values.
Asymmetry in the  conductance change, i.e., the average increment (potentiation) and decrement (depression)
    in conductance that can be reliably realized in a device is also an important challenge.  Some devices also show significant state dependence where the conductance
update depends on the current state of the device. For instance, this makes potentiation progressively harder with increasing conductance values. We
refer to this as non-linear conductance response. In addition to weight update challenges,
random volatile conductance fluctuations in the constituent elements of the computational memory and the finite resolution of the data converters used to interface them with the processing units could significantly impact the accuracy of the
computations performed. In this paper, we describe the mixed-precision architecture based on the computational memory and describe how it can address the aforementioned challenges. We use a neural network meant for classifying handwritten digits to benchmark the system performance.

\section{Mixed-precision architecture based on computational memory}

In supervised training of DNNs, the weights of the network are optimized based on a training data set. The computations during
training can be divided into three main stages; forward propagation, backward propagation, and weight update. During the forward propagation, an
instance from the training data set is presented to the input layer and any subsequent layer will receive a weighted sum of the outputs from all or a
subset of the neurons in the previous layer. Typically a non-linear neuronal function (eg. sigmoid, tanh, ReLU etc.) is applied over this weighted sum and is
propagated to the next layer. The last layer neurons' response is compared with the dataset label and an objective function based on this observed
and desired network response is minimized by altering the synaptic weights using a gradient descent algorithm. To train weights in the inner layers
of the network, the gradient of the objective function with respect to the weights from the final layer needs to be back propagated based on the chain rule in differentiation.  This back-propagation
involves the weighted sum of an error signal from the previous layer neurons.  Finally, the synaptic weight update can be determined as a product of
the back-propagated errors and neuron  activations. This process is repeated several times over multiple training examples to arrive at a weight
distribution that enable the network to provide a satisfactory classification/detection accuracy.

\begin{figure}[h!]
\centering
\begin{tabular}{c}
\includegraphics[width =0.95 \columnwidth]{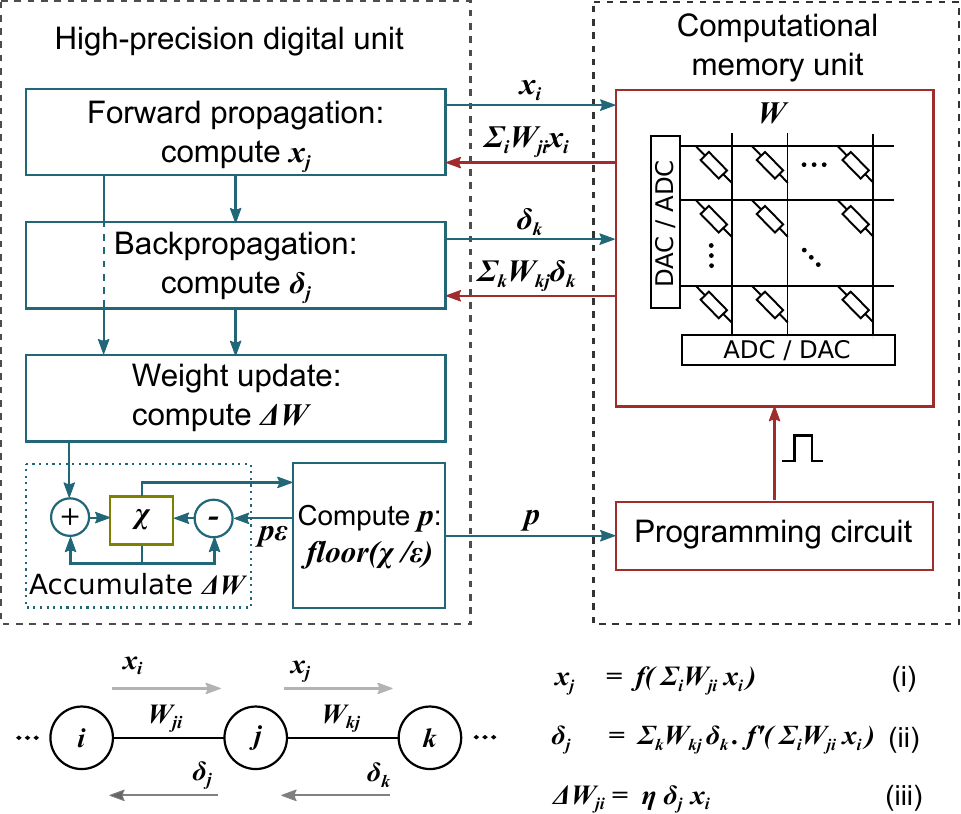}
\end{tabular}
\caption{\textbf{Mixed-precision architecture based on computational memory:} The synaptic weights are stored in a computational memory unit as
conductance states of resistive memory devices organized in crossbar arrays. The matrix-vector multiplications associated with the forward and the
backward propagation are performed in place in the memory arrays. The weight updates are accumulated in a volatile memory, $\chi$ in high precision
until they become comparable to the update granularity ($\epsilon$) of the memory devices. The device updates are integer multiples of $\epsilon$ and
the same quantity will be subtracted from $\chi$.} \label{fig:scheme}
\end{figure}

In Fig. \ref{fig:scheme}, we introduce the mixed-precision computational memory approach for training DNNs. The most expensive
operation during the forward and backward propagation is determining the weighted sum, which are matrix-vector multiplications. A computational
memory unit which has resistive memory devices organized in a crossbar array is ideally suited to perform these matrix-vector operations with
constant computational time complexity.\cite{Y2016huDAC} The neuron activations of a layer, $x_i$, are applied as voltages to the word lines using digital-to-analog
converters (DACs). Currents proportional to the conductances will flow through the devices,  and the resulting total current flowing through any bit
line will be $I_j= \Sigma _i W_{ji}x_i$. Here $W_{ji}$ represent the device conductance connecting neuron $i$ to a next-layer neuron $j$. These
currents read and digitized using analog-to-digital converters (ADCs) will be the desired weighted sum operation results. The same crossbar array can
be shared to perform the matrix multiplication during the back-propagation in the same layer. The errors to be back-propagated, $\delta_k$, are applied
as voltages to the bit lines and the currents are read out from the word lines, realizing a transposed matrix multiplication ($\Sigma_k
W_{kj}\delta_k$).

The desired weight updates are determined as the product of the back-propagated error and the neuron activation, $\Delta W_{ji} = \eta \delta_j x_i$,
where $\eta$ is the learning rate. Even though the computational memory unit can accelerate the forward and the backward propagation significantly,
updating the synaptic weights with the desired precision is very challenging. Often, the devices representing the synaptic weights have a conductance
update granularity dictated by the physical mechanism behind it. Let $\epsilon$ be the absolute value of the smallest conductance change that can be
reliably achieved. Attempts to program weight updates which are much smaller could induce significant error in the training. In the proposed approach, the weight updates are accumulated in high precision in a variable $\chi$. The device conductance will
be updated only if the magnitude of the accumulated weight update becomes greater than or equal to an integer multiple of $\epsilon$. The number of
programming pulses, $p$, to be applied to the resistive memory device is determined by flooring $\chi/\epsilon$ toward zero, and the same number of
$\epsilon$s is subtracted from the $\chi$. Depending on the sign of $p$, the conductance value of the corresponding device will be increased or
decreased. Note that, the actual conductance state of the devices are never read back, and hence we will never be able to confirm whether the requested
weight updates are accurately attained as equivalent conductance changes in the devices. In spite of this, we show that this scheme works
remarkably well and that the performance is often comparable to those of floating-point implementations. This single-shot programming method, which avoids
verification and iterative programming steps, enables the acceleration of the training process.  In subsequent sections, we will present a detailed evaluation of
this methodology under various scenarios of device-level non-ideal behavior.

\section{Evaluation of the mixed-precision architecture}
\subsection{The simulation framework}

\begin{figure}[h!]
\centering
\begin{tabular}{c}
\includegraphics[width = 0.95\columnwidth]{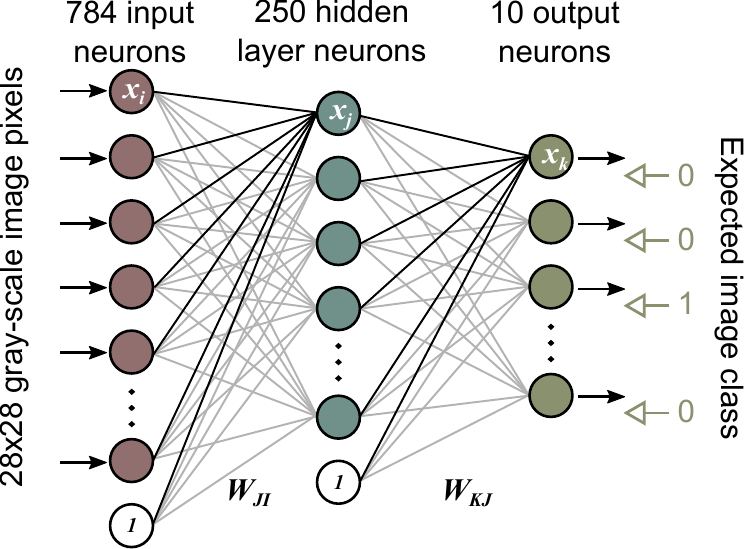}
\end{tabular}
\caption{\textbf{Neural network for digit classification:} The neural network used to evaluate the mixed-precision architecture. The
objective is handwritten digit classification based on the MNIST data set. There are 784  input neurons, 250 hidden sigmoid neurons, and 10
output sigmoid neurons. The network weights are trained by optimizing a quadratic objective function using gradient descent. All the 60,000 images in the dataset are used in one epoch of training and 10,000 images for testing.} \label{fig:mnist}
\end{figure}

The performance of the mixed-precision architecture is analyzed based on its classification accuracy on the MNIST handwritten digit dataset using a
neural network as shown schematically in Fig. \ref{fig:mnist}. The number of neurons in the input, the hidden and the output layer is 784, 250, and 10
respectively. The hidden and output neurons are sigmoid.  The network is trained using the entire training set of 60,000 images for ten epochs, and a
test accuracy is reported based on 10,000 test images. The pixel values of the $28\times 28$ gray-scale images are normalized between 0 and 1 before they are supplied as input to the
network. No other preprocessing is performed on the images. We used the quadratic objective function for the back-propagation-based training and used
a fixed learning rate. The network gave $98\,\%$ floating point (64-bit) test accuracy when trained using stochastic gradient descent. This
classification result is used as \textit{reference} to evaluate the performance of our mixed-precision approach.  The final weight distribution from
this high-precision training was approximately in the range [-1 1].

\subsection{Inaccuracies arising from weight-updates}
In this section, we will evaluate how the proposed architecture copes with the issues associated with weight updates. We assume a hypothetical linear
device with a conductance range of [-1 1] similar to the floating-point trained weight distribution. The device is assumed to have $n$-bit update granularity such that it covers its conductance range in  $2^{n}-2$ steps and hence it will have  $2^{n}-1$ possible levels in the absence of conductance change stochasticity. An odd number of levels was chosen to include zero. Therefore, in our mixed-precision training approach, we will update the device when the weight-update accumulation exceed the conductance change step size, $\epsilon = 2/(2^n-2)$.  In subsequent discussions, we will use both $\epsilon$ and $n$ interchangeably to indicate the granularity
associated with the weight-updates.

\begin{figure}[h!]
\centering
\begin{tabular}{c}
\includegraphics[width = \columnwidth]{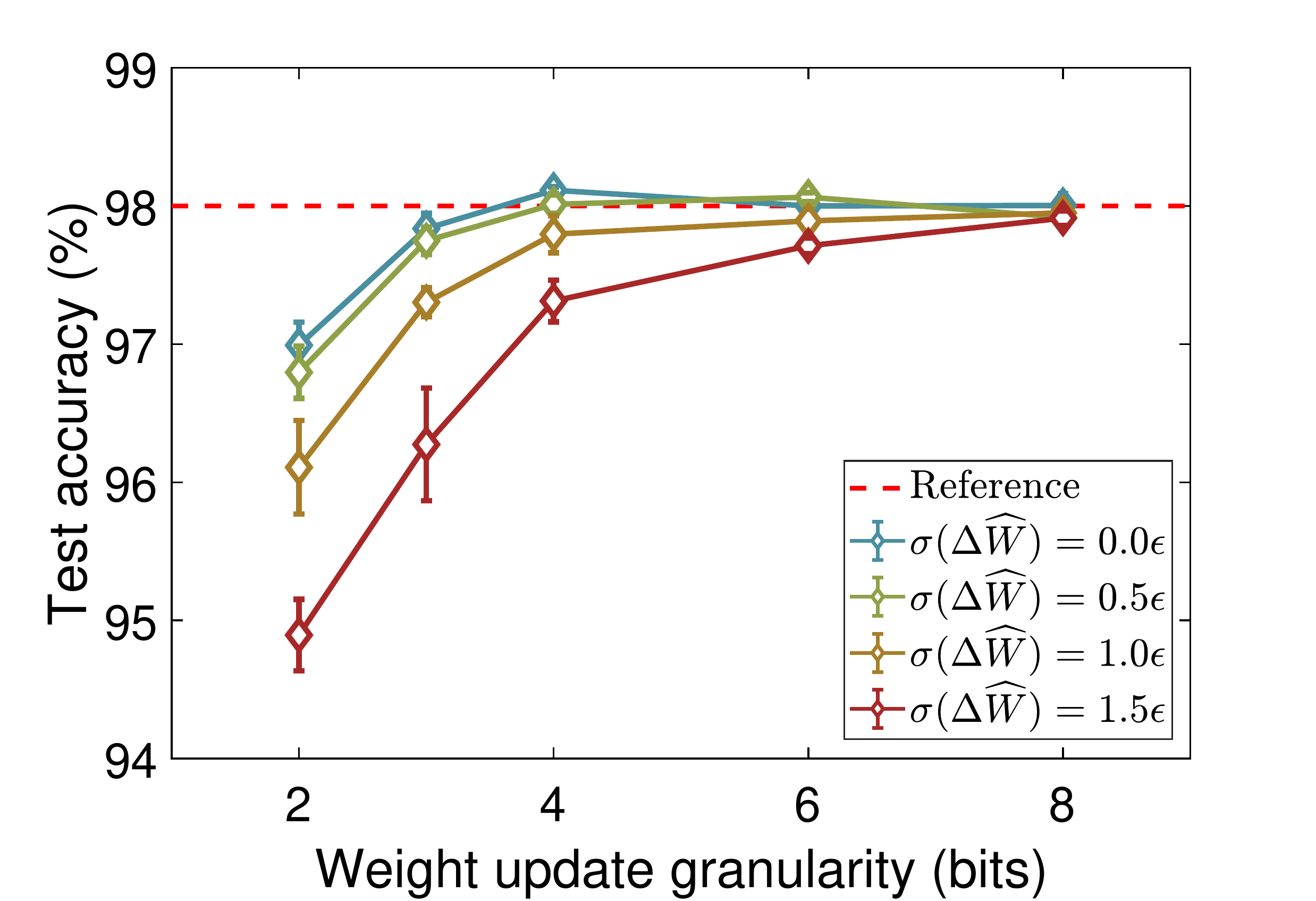}
\end{tabular}
\caption{\textbf{Effect of granularity and stochasticity associated with weight-updates:} Linear devices with symmetric potentiation and depression
granularity are assumed as computational memory elements. The standard deviation of the weight-update randomness, $\sigma(\Delta \widehat{W})$, is taken as a multiple of the weight-update
granularity, $\epsilon$. The error-bars indicate the standard deviation corresponding to five repetitions of the simulation. It can be seen that even
in the extreme cases of highly coarse and random weight-updates, drop in the  test accuracy is still within approximately $4\%$ for 2-bit granularity.}
\label{fig:stochastic}
\end{figure}

\begin{figure}[h!]
\centering
\begin{tabular}{c}
\includegraphics[width=\columnwidth]{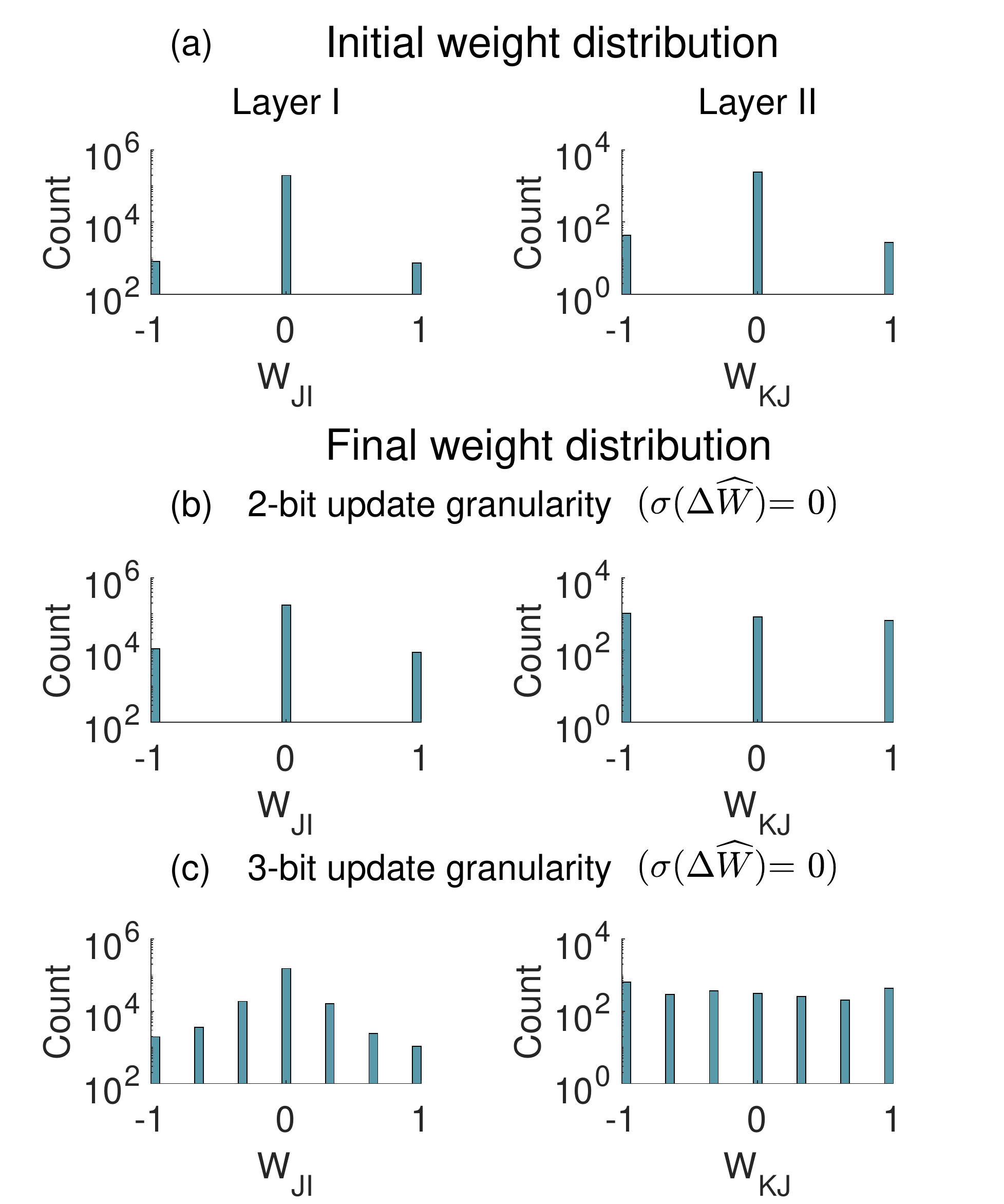}
\end{tabular}
\caption{\textbf{Discrete weight solutions:} (a) In the linear device simulations, the weights are initialized to a set of states  -1, 0, and
1. (b) In non-stochastic two bit granular updates the devices go through only these discrete states and hence the update granularity
also becomes the device resolution. This discrete weight solution gave a test accuracy of approximately 97\%.  However, in case of
stochastic programming, the devices can achieve intermediate states. (c) The final weight distribution from the 3-bit update granularity simulation is
also shown. The higher weight resolution improved the test accuracy by approximately 1\%.} \label{fig:Discretedistribution}
\end{figure}

The  conductance updates in the non-volatile devices are often stochastic. Even though it is desirable to induce a change in conductance corresponding to an integer multiple of $\epsilon$, the observed change is often quite different from the desired one. Therefore, the actual weight
update from the device, denoted by $\Delta \widehat{W}$, is modeled as a Gaussian random variable whose mean is $\epsilon$ and whose standard deviation
($\sigma$) is a fractional multiple of $\epsilon$.  This device model is used as the computational memory elements representing the neural network synapses during its training using the mixed precision scheme.  The devices are initialized to \{-1, 0, 1\} states with a discrete distribution whose variance is normalized by the
number of neurons in the pre- and post-synaptic layers. Device read noise and analog-digital converters are ignored at this stage.  The simulated
classification accuracies with limited granularity and with different amounts of stochasticity in the updates is shown in Fig. \ref{fig:stochastic}. In the case where the weight
updates are non-stochastic the test accuracy drop is only $~1\%$ for 2-bit, and with 3-bit granularity the accuracy is very close to that obtained
in the floating-point simulation (\textit{reference}). As the  stochasticity  increases, the performance degrades with reducing number of
bits. However, it is remarkable that even though the standard deviation of the weight update is equal to or greater than the mean weight update
granularity itself, drop in the test accuracy is still within approximately $4\%$ for 2-bit granularity. The test accuracy becomes closer to the reference floating-point accuracy as the device granularity is further reduced.

The distributions of the initial and trained weights in the two layers of the neural network  for the 2-bit and 3-bit update granularity are shown in
Fig. \ref{fig:Discretedistribution}. The distributions are shown for  non-stochastic device programming and hence the final weights are also discrete
and the number of levels correspond to the update granularity. We observe  that increasing the number of levels improved the classification
performance until an update granularity of 4-bit beyond which the test accuracies remained approximately constant. The stochasticity associated with
conductance updates helps to create a non-discrete weight distribution. However, we found this to have no significant advantage and we typically observe
a decrease in the classification performance with increasing stochasticity (Fig. \ref{fig:stochastic}).

\begin{figure}[t]
\centering
\begin{tabular}{c}
\includegraphics[width=\columnwidth]{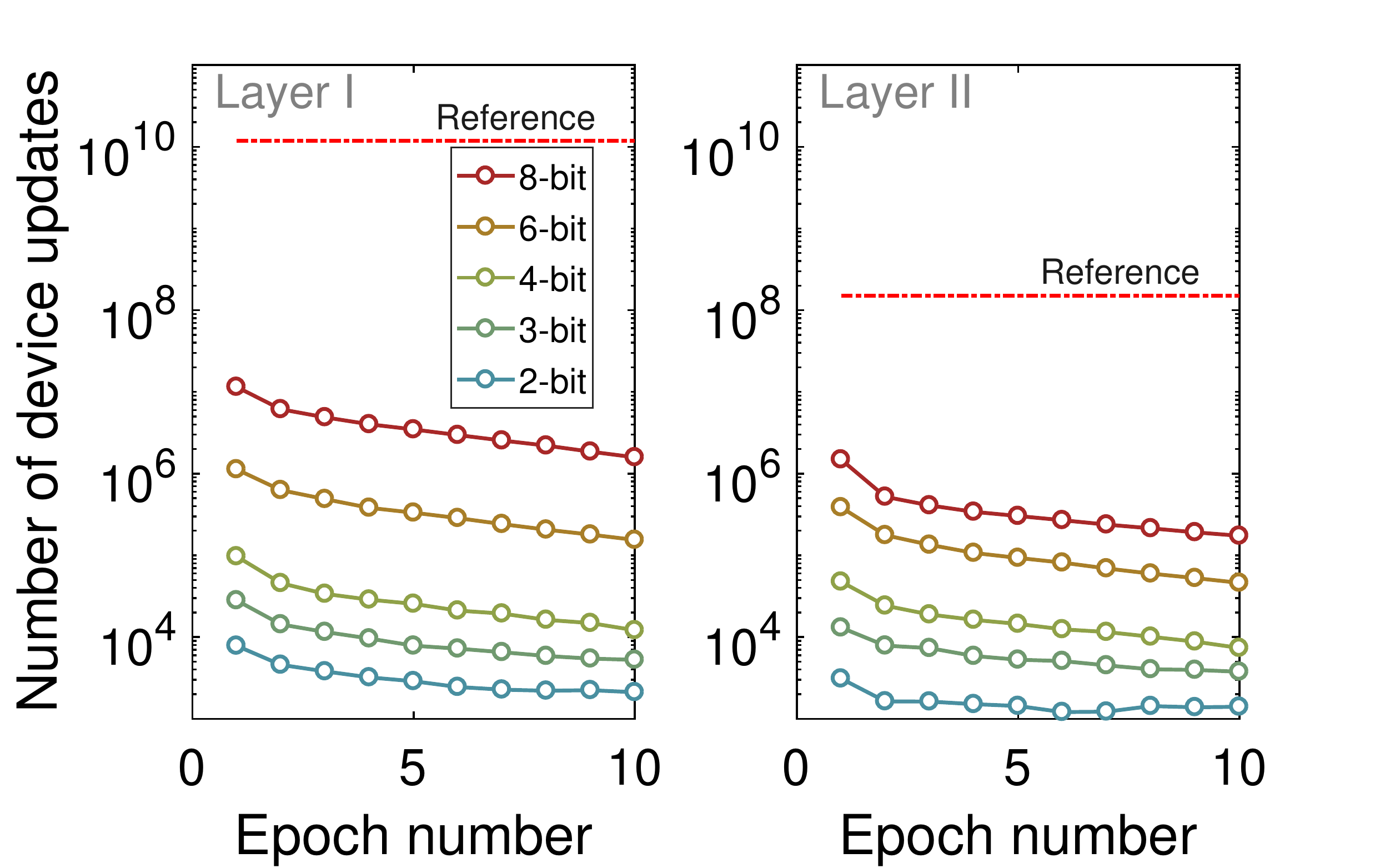}
\end{tabular}
\caption{\textbf{Sparsity associated with the devices that are being programed in the synaptic array:} The device update count per epoch is plotted for different values of
$\epsilon$. Only a fraction of the total number of devices is programmed after each image. The number of synapses in each layer times the total training image count is indicated as reference.  As the
value of  $\epsilon$ increases, the number of updates reduces by several orders of magnitude,  saving significant programming overhead.} \label{fig:NumDeviceUpdates}
\end{figure}

From the previous discussion, it can be seen that increasing the resolution of conductance change beyond a certain value does not necessarily improve
the network performance. Moreover, in the mixed precision scheme, there is a significant reduction in the device programming cost with the use of
larger $\epsilon$s. In Fig. \ref{fig:NumDeviceUpdates},  we show the number of device updates during each epoch of training. The maximum
number of device updates, calculated as the product of the synapse count and the training image count, assuming all
    the weights are updated after each image presentation, is indicated as reference. However, in the mixed precision approach, we accumulate the updates in high precision. As a
result, the smaller updates are combined and delivered together to the device. Hence, as the device update granularity ($\epsilon$) increases, the
 devices need to be programmed less often, resulting in eventual energy savings. Programming resistive memory
devices incurs significant time and power penalty and hence it is desirable to reduce the number of such programming instances, without
compromising the network performance.  For the chosen network architecture and classification problem, 4-bit update granularity seems to offer the
best case scenario of highest test accuracy with reduced programming expense.

\begin{figure}[h!]
\centering
\begin{tabular}{c}
\includegraphics[width = \columnwidth]{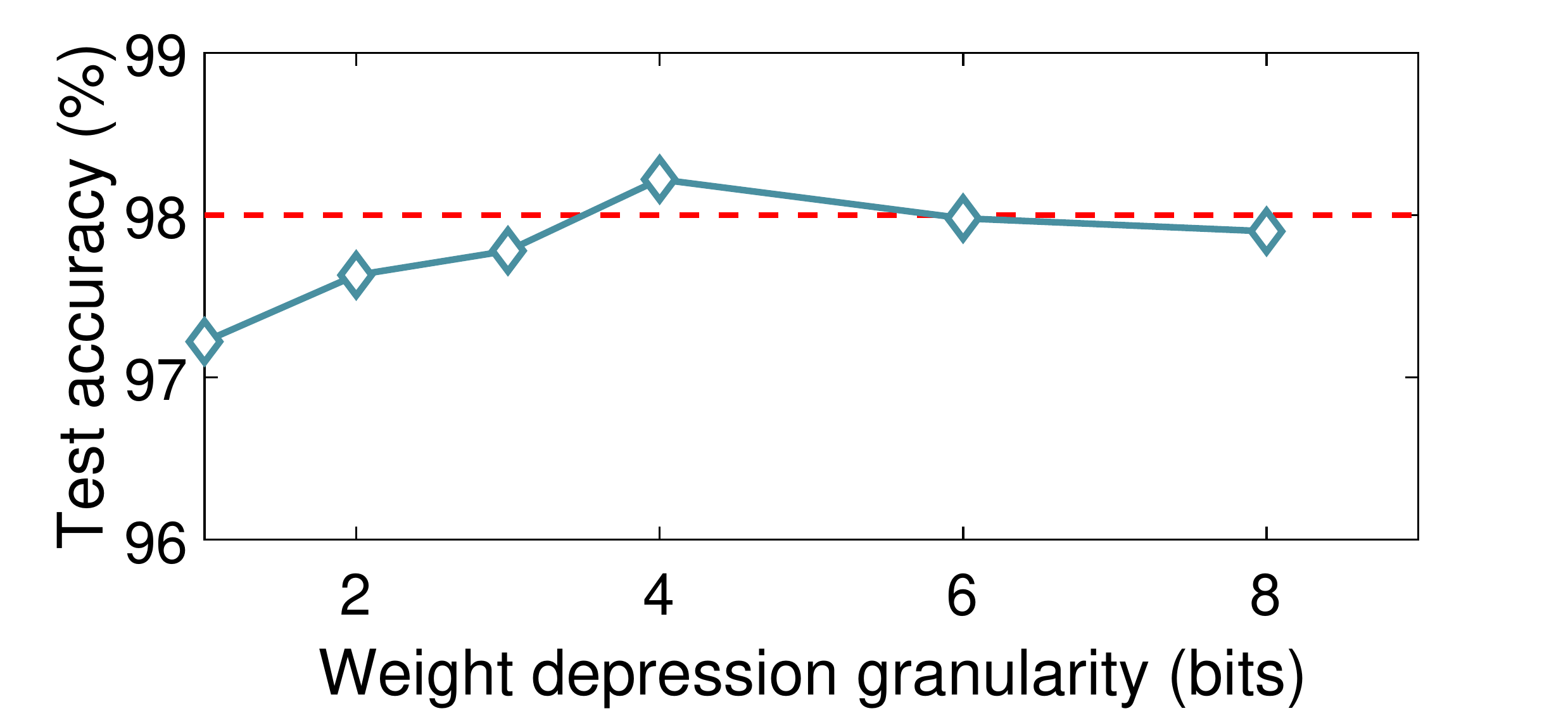}
\end{tabular}
\caption{\textbf{Effect of asymmetric conductance response:} The test accuracy, when trained with devices of fixed 8-bit potentiation granularity and
variable depression granularity, is plotted as a function of the depression granularity (expressed in bits). Weight updates are assumed to be
deterministic. The resulting test accuracy shows less than 1\% accuracy drop even in the highest asymmetric case in the simulation.}
\label{fig:Asymmetry}
\end{figure}

Next, we study the influence of asymmetric conductance update response. We assume a device with fixed but unequal potentiation and depression
granularity. The mixed-precision method can cope with this behavior by using different thresholds, $\epsilon_P$ for conductance increment and
$\epsilon_D$ for conductance decrement. For example, in Fig. \ref{fig:Asymmetry} we  assume an 8-bit potentiation granularity and the depression
granularity is varied. The one bit depression in the figure correspond to a situation where the update granularity, $\epsilon _D$,  equals the entire
weight range in contrast to the previous definition. The weight updates are assumed to be deterministic. The resulting test accuracies  show less
than 1\%  drop even for the maximum asymmetric case tested, demonstrating the efficacy of the proposed scheme to tolerate device update asymmetry effectively.

\begin{figure}[h!]
\centering
\begin{tabular}{c}
\includegraphics[width = \columnwidth]{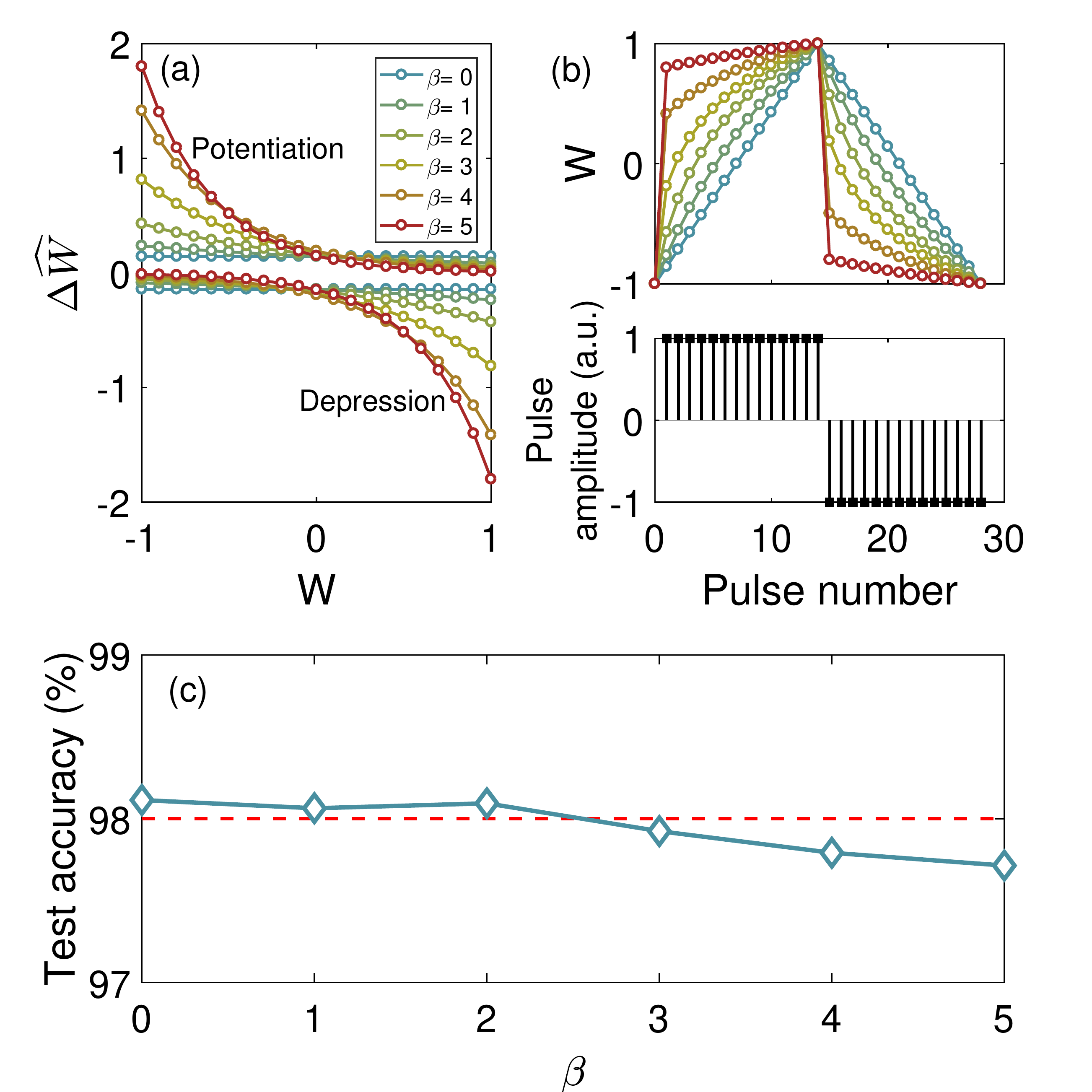}
\end{tabular}
\caption{\textbf{Effect of non-linear conductance response:} (a) Non-linear device model: the weight update ($\Delta W$) is modeled as exponentially dependent on the current state, $W$.
The exponential function for different amount of non-linearity is plotted. (b) Corresponding device model  pulse response, where 14
potentiation pulses followed by the same number of depression pulses are applied to the device. (c) Non-linear device model as synapse for DNN training. $\beta = 0$ correspond to a symmetric linear device and higher $\beta$ values indicate increasing amount of
non-linearity. Approximately 4-bit weight update granularity is assumed for the device model and mixed precision training. Weight updates were non-stochastic.  The result shows that there is no significant degradation in the test accuracy even for $\beta = 5$ that corresponds to a highly
non-linear conductance response.} \label{fig:Nonlinear}
\end{figure}

Subsequently, we investigate the influence of non-linear conductance response. To analyze this we simulated the training problem using a device model
whose non-linearity could be tuned. We chose an exponential function to model the state dependency of $\Delta W$ as suggested by Querlioz et al
\cite{Y2011querliozIJCNN}. The model essentially captures the behavior where a resistive memory device closer to the boundary conductance will exhibit smaller
update compared to those away from it, when updated towards the boundary.
\begin{equation}
\Delta \widehat{W}_{P} = \alpha e^{-\beta \frac{W-W_{min}}{W_{max}-W_{min}}}
\end{equation}
\begin{equation}
\Delta \widehat{W}_{D} = \alpha e^{-\beta \frac{W_{max}-W}{W_{max}-W_{min}}}
\end{equation}
Here, $\Delta  \widehat{W}_{P} $ and $\Delta  \widehat{W}_{D} $  model the potentiation and depression respectively for a device at a conductance of
$W$. $W_{min}$ and $W_{max}$ represent the limits of the device conductance. We used the parameter  $\beta$  to tune the amount of non-linearity and
$\alpha$ to adjust the update granularity. To make a reasonable comparison in training performance using models of different amount of non-linearity,
we assume that two criteria have to be satisfied:  the device models must have the same on-off ratio and they must take the same number of programming steps to span
the whole conductance range, irrespective of the non-linearity. The $\Delta  \widehat{W}$ versus $W$, and $W$ versus pulse number responses
satisfying these conditions for different values of $\beta$ are shown in \ref{fig:Nonlinear}(a) and (b). Here, $\beta = 0$ correspond to a linear
device. The number of pulses for full range potentiation or depression is assumed to be corresponding to that of a 4-bit update granularity. The
same update granularity is assumed to determine the $\epsilon$ for the mixed precision scheme for varying amount of non-linearity.  The resulting
test accuracies are plotted as a function of $\beta$ in Fig. \ref{fig:Nonlinear}(c). It can be seen that there is no significant degradation in the
test accuracy even for $\beta = 5$, which is very close to  the behavior of a binary device.

\subsection{Inaccuracies arising from matrix-vector multiplication}
\begin{figure}[h!]
\centering
\begin{tabular}{c}
\includegraphics[width = \columnwidth]{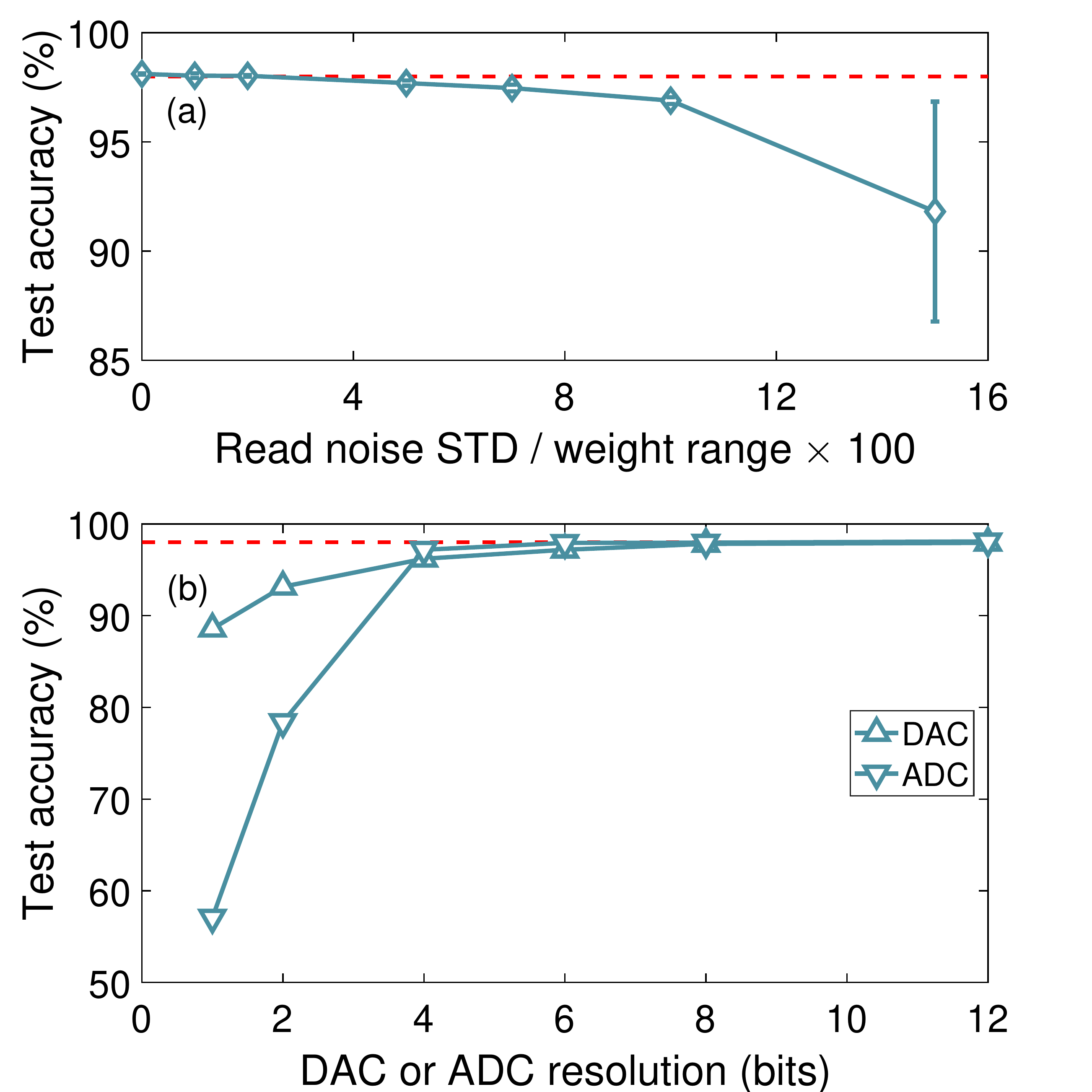}
\end{tabular}
\caption{\textbf{Matrix-vector multiplication errors:} (a) Effect of read noise. Gaussian distributed additive read noise is added with the computational memory devices whenever they are used for multiplication. The standard deviation (STD) of the read noise is varied as a fraction of the total     device conductance  range. For the mixed precision training, a 4-bit weight update granularity and non-stochastic programming is assumed. There is no significant loss in test accuracy even up to a read noise corresponding to 5\% of the total weight range.  (b) Effect of
finite resolution data converter. The weight update granularity is assumed to be 4-bit, without stochasticity and read noise.The curve with triangle indicates
simulation results where DACs are used at the crossbar input whereas the
output current is read back in floating-point precision. The curve with inverted
triangle indicates results where the crossbar input has floating point precision
whereas ADCs are used for reading back the output current.}
 \label{fig:readnoise}
 \label{fig:DACADC}
\end{figure}

In this section, we analyze the influence of conductance fluctuations and finite resolution of data converters. Resistive memory devices typically
exhibit fluctuations in conductance arising from trapping/detrapping processes\cite{fugazza2009}. The effect of this read noise in the DNN training
using the proposed scheme is tested by adding a zero mean white Gaussian noise to a linear device model. The noise is added  to the weights whenever
it is used in the matrix multiplication in the forward  and the backward propagation. It is also incorporated during the testing phase (only forward
propagation). The standard deviation of the noise is varied as a fraction of the total weight range.  The resulting test accuracies are shown in Fig.
\ref{fig:readnoise}a. It can seen that the methodology is quite robust to up to $5\%$ read noise.

An additional source of error in the matrix-vector multiplication is due to quantization from the DACs and ADCs. During forward propagation, the
neuron activations evaluated in the digital domain are converted to analog voltages using DACs before they are applied to the word lines of the
crossbar array.  The weighted sum obtained as currents in the bit lines are read back using ADCs. Similarly, the back-propagated errors are converted
to analog voltages when applied to the cross-bar array matrices. The range for the digital to analog converters are fixed for sigmoid and tanh neuron
activations, whereas for the ReLu neurons this could be a challenge as their range dependents on the data and weight distribution. Here, we chose
sigmoid neurons for our network, which fixed the DAC range in  the forward propagation. Furthermore, we normalized the back-propagated errors to fix
the range for its interface converters to analog voltages. The normalization factor is multiplied with the learning rate during the weight update
calculation. However, the input for the analog to digital converters are results from matrix-vector multiplications and their distribution is dependent on the number of neurons and the weight distribution in the layer. In this work, the range for ADC was determined by observing the distribution of corresponding variables representing the weighted sums.  To study the effect of
DACs and ADCs separately, the bit precision of one of them is varied, whereas the other variables are
represented in floating-point precision. Fig. \ref{fig:DACADC}b shows that  8-bit resolution is sufficient to avoid a noticeable degradation in test accuracy.

\subsection{Phase-change memory synapses}
Phase-change memory  is a relatively mature resistive memory technology that has found applications in the space of storage-class memory and novel
computing paradigms such as neuromorphic computing \cite{Y2016tumaNatNano,Y2016tumaEDL,pantazi2016}  and computational memory
\cite{Y2017sebastianNatComm,Y2017legalloArXiv,Y2017legalloIEDM}. It is based on the property of chalcogenide alloys, typically compounds of Ge,
Sb and Te, which exhibit drastically different electrical resistivity depending on whether they are in the ordered crystalline phase or in the
disordered amorphous phase. The crystalline phase of the Ge$_2$Sb$_2$Te$_5$ (GST) alloy is orders of magnitude more conductive than the amorphous phase. If this material is sandwiched
between two metal electrodes, the phase-configuration of the material and thus the
    conductance of the device can be reversibly changed by applying suitable electrical pulses.  The crystalline to amorphous phase transition is accomplished by a melt-quench process and the reverse transition is
governed by temperature accelerated nucleation and crystal  growth \cite{Y2014sebastianNatComm}. It is possible to achieve a continuum of conductance values
by partial crystallization or amorphization in these devices. This analog storage capability makes PCM particularly suited for applications in the
space of computational memory \cite{Y2017legalloArXiv,Y2017legalloIEDM}.

\begin{figure}[h!]
    \centering
    \begin{tabular}{c}
        \includegraphics[width = \columnwidth]{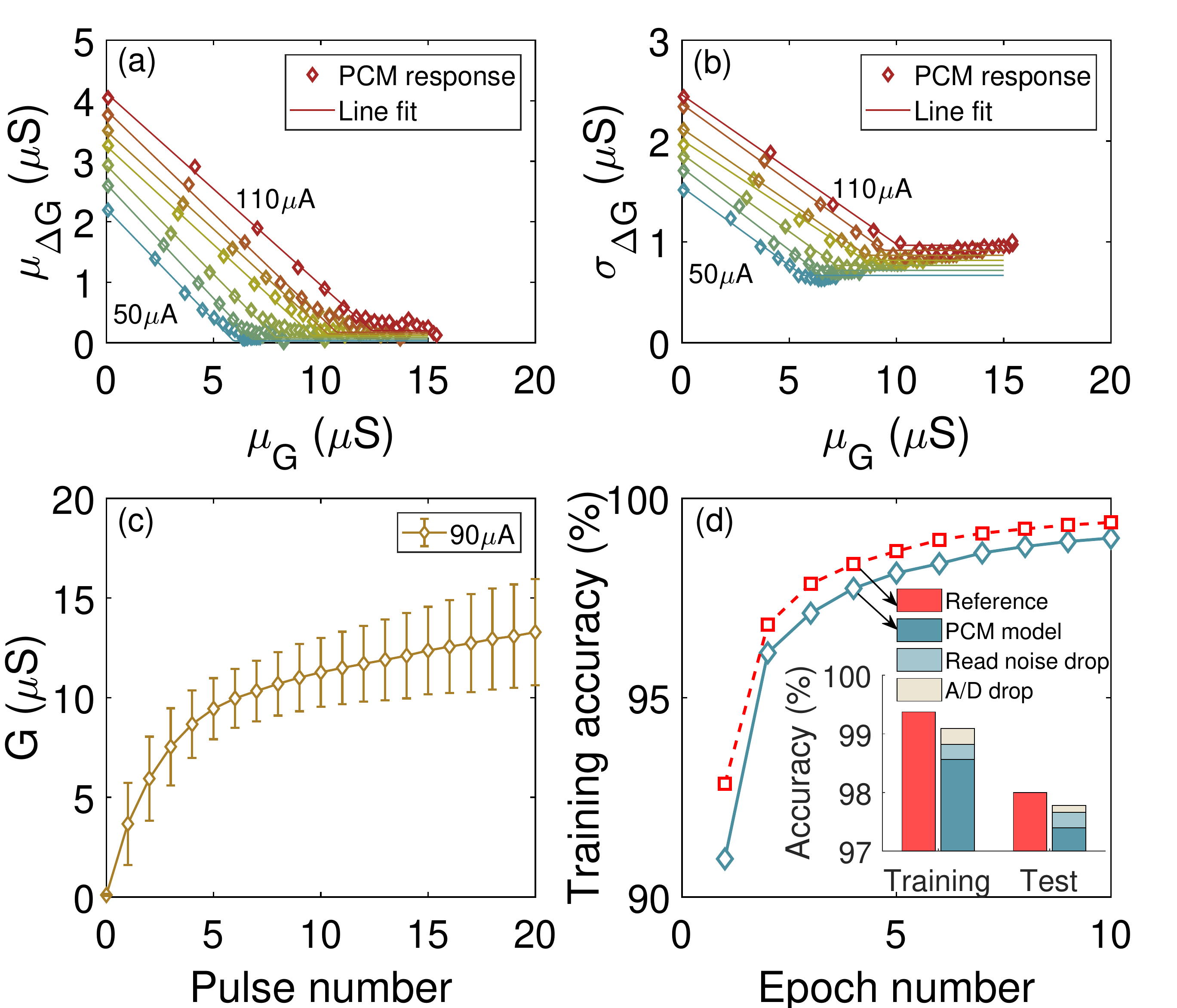}
    \end{tabular}
    \caption{\textbf{Training using PCM synapse models:}  Piece-wise linear approximations to (a) the
   mean ($\mu$) and (b) the standard deviation ($\sigma$) of the experimentally measured
   $\Delta G$ for Ge$_2$Sb$_2$Te$_5$-based PCMs for their average current conductance state
   ($\mu_G$) are used to model the device. (c) The resulting model conductance evolution
   in cumulative pulse programming. (d) Training using PCM
   models. Two non-linear PCM device models in differential configuration are
   used at the cross-points for the neural network weights. Training convergence
   and test accuracies (inset) are shown. Device-model based network simulation
   achieves 97.78\% test accuracy. Additional drop from the read noise (0.26\%)
   and analog-digital converters (0.12\%) are indicated.}
    \label{fig:PCMsim}
    \label{fig:PCMmodel}
\end{figure}

The PCM devices exhibit most of the non-idealities we described earlier, such as limited granularity, non-linear and asymmetric conductance
update, and programming stochasticity. There is also substantial read noise associated with these devices. To evaluate the suitability of PCM devices
for the mixed-precision approach to train DNNs, we developed a model that captures the essential physical attributes of PCM
devices. The model is created based on characterization data from approximately 10,000 devices integrated in $90\,$nm CMOS technology
\cite{Y2017nandakumarDRC}. The devices are subjected to 20 programming pulses of fixed amplitude and each state is read 50 times to eliminate read noise. The mean and standard deviation of the extracted conductance change ($\Delta G$) versus the average initial conductance for each
programming pulse are fitted using piece-wise linear models as shown in the Fig. \ref{fig:PCMmodel}a,b. Assuming the $\Delta G$ to be a Gaussian random
variable, the device cumulative pulse response is simulated, and the statistical plot of the resulting stochastic model behavior is plotted in Fig.
\ref{fig:PCMmodel}c.

This device model was used to emulate the synapses in the crossbar array to study its influence on training DNNs. Two PCM devices in
differential configuration with weight refresh \cite{Y2015burrTED,suri2011} are used. The conductance values are initialized to a normal distribution
around $2\,\mu$S whose standard deviation is normalized based on the number of neurons in the pre- and post-synaptic layers. Resulting test accuracy
after 10 epochs of training was 97.78\% (Fig. \ref{fig:PCMsim}d). Incorporating a fixed read noise (zero mean Gaussian noise with experimentally
measured average standard deviation) and 8-bit analog-digital converters during training and testing resulted in an additional 0.38\% drop in
accuracy. We also tested the training performance where each synapse is realized using a single PCM device model at the crosspoint, exploiting the
capability of the scheme to cope with the strongly asymmetric conductance response. The final test accuracy for the MNIST dataset classification was
96.5\%, indicating the robustness of our scheme.

\section{Discussion}

The non-volatile memory crossbar array based computational memory unit is ideally suited to perform matrix-vector multiplications. By utilizing the
computational memory to perform those operations when training DNNs, the forward and the backward propagation of data can be
significantly accelerated. Also, the processor-memory bottleneck is reduced as the synaptic weights are not transferred during the propagations. However,
the necessity to frequently update the memory devices poses an additional challenge compared to applications of computational memory where the
matrix does not change \cite{Y2017legalloIEDM,Y2017legalloArXiv}. In back-propagation based training algorithm it is desirable to update the weight
matrix after the presentation of each training instances. Using devices like PCM, which can attain a continuum of conductance states, it is
possible to iteratively program the devices to the desired conductance states accurately\cite{Y2011papandreouISCAS}. However, this involves repeated
read/write cycles and incur significant time/power penalty. The necessity to program a large number of devices very often could overshadow the
performance gain that we obtain from the in-place matrix-vector multiplication in the crossbar array.

On the other end of the spectrum lies the non-von Neumann coprocessor approach proposed recently
\cite{Y2015burrTED,Y2016gokmenFN,Y2017kimArXiv}. As before, the synaptic weights are stored in resistive memory devices organized in crossbar arrays
and the  matrix-vector multiplications during forward and backward propagation are realized in place using these arrays. However, they suggest a
fully parallel conductance update by overlapping pulses from the pre- and post-synaptic neuron layers. By realizing the neurons and associated
circuits in place, this offers the possibility of a fully parallel non-von Neumann system. By accelerating all the three components of training DNNs,
namely forward propagation, backward propagation, and weight update, this approach could be the fastest and most energy-efficient compared to alternate approaches. However,
the non-idealities associated with programming the memory devices will pose significant challenges in realizing state-of the art classification
accuracies. An ideal device is expected to have a symmetric weight update granularity of 10 bits \cite{Y2016gokmenFN}. Experimental demonstrations
using more realistic phase change memory devices have shown a limited test accuracy of less than $83\%$\cite{Y2015burrTED}.

Our mixed precision approach is designed to take into account the limited device update granularity seen in experimental devices today. The proposed architecture is  significantly
tolerant to conductance programming asymmetry and update non-linearity. In contrast to the above discussed methods, we deliver the conductance
updates only when weight updates accumulated in high precision become comparable to the device update size. As a result, the number of device
programming instances are reduced by several orders of magnitude as the update size increases. As a result, the advantage of matrix-vector multiplication
acceleration in the data propagation stages is preserved  without significant device programming overhead. We follow a blind single pulse programming approach
without read-back to deliver an $\epsilon$ amount of update. The value of $\epsilon$  is chosen based on the  device dynamics. The simulations show
that the resulting sparse weight updates training are able to achieve classification accuracies comparable  to those from the floating-point
simulations in similar number of training epochs. Further, the high precision accumulation and less frequent weight-updates combined with the
inherent error tolerance of neural network training enable the architecture to cope with the high device programming stochasticity.

We believe that the weight update and accumulation overhead associated with this mixed precision architecture is significantly less compared to the
training acceleration we obtain. The training acceleration is achieved by computing the  multiply-accumulate operation of approximately   $\mathcal{O}(N^2)$
complexity in  fixed time for each $N\times N$ neural network layer. The device updates are sparse and the weight update accumulation in high
precision is equivalent to the weight update scheme in  standard stochastic gradient decent except that the memory is initialized to zero here. The
additional thresholding/flooring and subtraction operations are computationally simple and do not incur additional memory read/write operations as
they can be preformed concurrently with the weight-update accumulation. Still, it is desirable to further accelerate the weight update stage of DNN
training as the weight-update determination is an  $\mathcal{O}(N^2)$  operation.

\section{Conclusion}
In this work, we presented a mixed precision computing architecture to train deep neural networks. The essential idea is to use a computational
memory unit in conjunction with a high precision processing unit. The computational memory unit comprises of resistive memory devices that are
organized in a crossbar array. The synaptic weights are stored as conductance states of these memory devices. The computationally expensive
matrix-vector multiplications arising during the forward and backward propagation stages of the backpropogation algorithm can be realized in a highly
efficient manner using this computational memory unit. However, the weight updates are accumulated in high precision and are only sporadically
transferred to the device array. This mixed-precision approach is shown to overcome the non-ideal behavior related to resistive memory devices such
as limited granularity and stochasticity associated with their programming as well as asymmetric and non-linear conductance response. In spite of the
added complexity arising from the high precision unit, we still gain in overall performance due to the substantial gain in time/power efficiency
associated with the forward and backward propagation steps. Moreover, the weight updates are sparse enough to not incur a significant time/power
penalty arising from the need to program the memory devices. This approach was tested using the MNIST handwritten digit classification problem and is
shown to achieve remarkably high classification accuracies even with computational memory units comprising single phase-change memory devices.
Realistic models of PCM devices fabricated in $90\,$nm technology node were used for this evaluation.

\section{Acknowledgements}
We would like to acknowledge fruitful discussions with C. Bekas, C. Malossi, G. Mariani and T. Moraitis.

\section*{References}

\end{document}